\documentclass[12pt,preprint]{aastex}
 \begin{document}


\title{Torsional nodeless vibrations
of quaking neutron star restored by combined forces of shear elastic and magnetic field stresses}


\author{
S.I. Bastrukov\altaffilmark{1,3}, G.-T. Chen\altaffilmark{2}, H.-K. Chang\altaffilmark{1,2}, I.V. Molodtsova\altaffilmark{3}, and D.V. Podgainy\altaffilmark{3}}


\altaffiltext{1}{Institute of Astronomy,\\
  National Tsing Hua University, Hsinchu, 30013, Taiwan}

\altaffiltext{2}{Department of Physics,\\
  National Tsing Hua University, Hsinchu, 30013, Taiwan}

\altaffiltext{3}{Laboratory of Informational Technologies,\\
Joint Institute for Nuclear Research, 141980 Dubna, Russia}


\begin{abstract}
 Within the framework of Newtonian magneto-solid-mechanics, relying on equations appropriate for a perfectly conducting elastic continuous medium threaded by a uniform magnetic field, the asteroseismic model of a neutron star
 undergoing axisymmetric global torsional nodeless vibrations under the combined action of Hooke's elastic and Lorentz magnetic forces is considered with emphasis on a toroidal Alfv\'en mode of
 differentially rotational vibrations about the dipole magnetic moment axis of the star. The
 obtained spectral equation for frequency is applied to $\ell$-pole identification of quasi-periodic oscillations (QPOs) of X-ray flux during the giant flares of SGR 1806-20 and SGR 1900+14.
 Our calculations suggest that detected QPOs can be consistently interpreted, within the framework of this model, as produced by global torsional nodeless
 vibrations of quaking magnetar if they are considered to be restored by the joint action of bulk forces of shear elastic and magnetic field stresses.
\end{abstract}

\section{Introduction}
 The recent detection of quasi-periodic oscillations (QPOs) on the light-curve tails of X-ray flaring SGR 1806-20 and SGR 1900+14 (Israel et al 2005; Watts \& Strohmayer 2006) that have been ascribed to torsional seismic vibrations of quaking magnetar raises several questions of quite general interest for the asteroseismology of degenerate solid stars.
  Namely, whether these differentially rotational oscillations are
 predominately of an elastic nature, that is, restored by Hooke's force of shear solid-mechanical stresses or they should be thought of as the toroidal  magnetic Alfv\'en mode of axisymmetric vibrations about dipole magnetic moment axis of the star and
 restored by the Lorentz force of magnetic field stresses (Glampedakis, Samuelsson \& Andersson 2006).
 Also, it remains questionable whether these vibrations are of global character, that is, excited in the entire volume or they can be
 explained as locked in the peripheral finite-depth crustal region of a neutron star.
 These and related issues are currently the subject of intense theoretical investigations
 by classical theory methods of material continua  (Piro 2005; Levin 2007, Lee 2008; Bastrukov et al, 2007a; 2008a) and by general relativity  methods (Sotani, Kokkotas \& Stergioulas 2008; Samuelssen \& Andersson 2007).

 In this article, continuing the above investigation, we study an asteroseismic
 model of a neutron star with uniform internal and dipolar external magnetic field undergoing quake-induced global differentially rotational, torsional, vibrations
 about the dipole magnetic moment axis of the star under the joint action of the elastic Hooke's and the magnetic Lorentz forces. In so doing we confine our attention to the regime of extremely long wavelength differentially rotational fluctuations of material displacements about motionless stationary state which are insensitive to core-crust compositional stratification of quaking neutron star.
 The characteristic feature of this regime is that conducting and highly robust to compressional distortions
 of solid-state material, of both core and crust, coupled by Maxwell stresses of uniform fossil magnetic field frozen in the star, sets in coherent axisymmetric differentially rotational vibrations
 with the nodeless toroidal field of material displacements identical to that for global torsional vibrations of a spherical mass of an elastic continuous medium capable of transmitting perturbation by transverse shear elastic waves generic to solid state of condensed matter, not a liquid one\footnote{It is commonly agreed that the average speed of transverse wave of elastic shear in the crustal matter is $c_t=\sqrt{\mu/\rho}\sim 10^8$ cm s$^{-1}$ which holds for both the outer crust with the density $10^{8}< \rho < 10^{11}$ g cm$^{-3}$ and
 the inner crust, whose density is ranged between the neutron drip line density  $\rho \sim 10^{11}$ g cm$^{-3}$ and density of overwhelming neutronization $\rho\sim 10^{13}$ g cm$^{-3}$ (e.g., Blaes et al 1990). Accordingly,
 the shear modulus of the crustal matter is ranged from the surface to the crust-core interface as $10^{24}< \mu < 10^{29}$ dyn cm$^{-2}$.
 The much denser core matter compressed by self-gravity to nearly normal nuclear density  $\rho=2.8\cdot 10^{14}$ g cm$^{-3}$ is much harder to elastic shear distortions and its shear modulus is estimated as $\mu \sim 10^{33}$ dyn cm$^{-2}$ (e.g., Owen 2005, Bastrukov et al 2008b and references therein); in view of this
 the Fermi-degenerate neutron matter of atomic nuclei and neutron star cores is designated as an elastic Fermi-solid capable of transmitting shear elastic wave with the speed $c_t=\sqrt{\mu/\rho}\sim 0.2\,c$ where
 $c$ being the speed of light.}.
 In the context of the above QPOs problem, such vibrations have been analyzed in some details in our recent works (Bastrukov et al 2007a, 2007b, 2008a), but the driving force was assumed to be of only elastic nature, that
 is, owing its origin to fluctuations in shear elastic stresses.
 The focus of this paper is placed, therefore, on the toroidal Alfv\'en vibrational mode in which magnetic field and field of material displacement undergo coupled fluctuations restored by Lorentz force with  Amp\'er's
 form of the conduction current density.
 Before proceeding to the details of calculations it worth noting that the liquid star model with axisymmetric poloidal uniform magnetic field inside and dipolar outside
 has been the subject of serious investigations in the past in the context of magnetic variables (e.g., Schwarzschild 1949; Chandrasekhar \& Fermi 1953; Ledoux \& Walraven 1958). The situation may be quite different for ultra strong
 internal magnetic fields frozen-in the super dense matter of the end products of stellar evolution, white dwarfs and neutron stars. In neutron stars self-gravity is counterbalanced by the degeneracy pressure of relativistic electrons in the crust, whose highly conducting matter is capable of sustaining persistent current-carrying flows and
 by degeneracy pressure of non-relativistic neutrons in the cores of neutron stars whose material
 can be in the state of paramagnetic magnetization caused by Pauli's mechanism of
 field-induced alignment of spin magnetic moment of neutrons along the frozen-in the star fossil magnetic field
 (e.g., Bastrukov et al; 2002a; 2002b). Of course, there is no compelling evidence that this is the case and there is no certainly general agreement how the magnetic fields of pulsars and magnetars are produced.
 Over the years a model of a neutron star with uniform internal and dipolar external field, pictured in Fig.1, has been invoked to discussions of evolution of magnetic dipole fields of isolated radio pulsars (Flowers \& Ruderman 1977), whose surface dipolar magnetic fields are found to be highly stable to spontaneous decay (Bhattacharya \& van den Heuvel 1991, Chanmugam 1994), as well as of origin and evolution of ultra strong magnetic fields of magnetars (Braithwaite \& Spruit 2006; Geppert \& Rheinhardt 2006; Spruit 2008), therefore the study of axisymmetric torsional vibrations within the framework of a neutron star model with uniform internal magnetic field seems amply justified.
 This interesting in its own right model describes situation when physically meaningful analytical solution of the eigenfrequency problem can be found and discloses mathematical difficulties one must confront when
 computing the frequency of Alfv\'en vibrational modes in the neutron star models with non-homogeneous internal and dipolar external magnetic fields. An analytic example of such latter field is discussed in (Roberts 1981; Geppert \& Rheinhardt 2006).

 The paper is organized as follows. In Sec.2, the exact solution is given for the
 eigenfrequency problem of global torsional nodeless vibrations of a solid star with a uniform field inside and dipole outside driven  solely by Lorentz restoring force. The obtained frequency spectrum of toroidal and poloidal magnetic Alfv\'en vibrational modes in such a star is then quantified numerically with input parameters, such as  mass, radius and magnetic field, characteristic to pulsars and magnetars. In Sec.3, the frequency spectrum of axisymmetric global torsional nodeless vibrations under the action of the combined forces of shear elastic and magnetic field stresses is obtained and applied to $\ell$-pole identification of the QPOs frequencies in flaring SGR 1806-20 and SGR 1900+14.
 The newly obtained results are briefly summarized in Sec.4.

\section{\bf Magnetic Alfv\'en modes of global nodeless vibrations
about axis of magnetic dipole moment of a neutron star}

  It is of common knowledge today that shortly before discovery of pulsars it has been
 realized that neutron stars must come into existence as compact sources of super strong dipolar magnetic field
 operating as a chief promoter of their electromagnetic activity.
 In particular, it has been suggested that in view of the absence in neutron stars of thermonuclear source of
 electromagnetic emission from the star surface they could manifest their presence in the universe by the conversion of energy of hydromagnetic Alfv\'en vibrations in a highly incompressible star mater into the energy of electromagnetic waves propagating out into the space (Hoyle, Narlikar \& Wheeler 1964). In this work we investigate this proposition in the context of current development of the neutron star asteroseismology by proceeding from
 equations of magneto-solid-mechanics underlying study of non-compressional Alfv\'en magnetic oscillations in
 quaking neutron stars.
 Such an approach implies that neutron star material, regarded as a perfectly conducting continuous medium threaded by a magnetic field frozen in the star on the final stage of gravitational collapse of its massive main sequence (MS) progenitor, possesses mechanical properties of super dense strained solid highly robust to compressional distortions, rather than a flowing liquid, as is the case of MS stars whose theoretical asteroseismology relies of equations of fluid-mechanics and magnetohydrodynamics. To this end, it worth emphasizing that contrary to the MS stars whose magnetic fields are generated by self-exciting dynamo process in a peripheral convective zone and the energy supply of this process
 comes from the central thermonuclear reactive zone (e.g. Parker 1979), in neutron stars there are no internal energy sources to drive relative motions of extremely dense conducting material which is under immense gravitational pressure so that mobility of neutron star matter is heavily
 suppressed. The common belief today is that magnetic fields of neutron stars are
 fossil (e.g. Fowlers \& Ruderman 1977, Bhattacharya \& van den Heuvel 1991, Chanmugam 1994; Spruit 2008), that
 is, they have been inherited from a massive MS progenitors and amplified in the processes of implosive gravitational collapse proceeding under control of the magnetic flux conservation. Also it is commonly agreed that
 external magnetic field of pulsars and magnetars is of dipolar symmetry, but the geometrical
 shape of the fossil magnetic magnetic field frozen-in the super dense motionless matter of the
 star remains fairly uncertain. The subject of present paper to formulate solid-mechanical
 variational scheme of computing frequency of Alfv\'en oscillations in the neutron stars with arbitrary
 configuration of frozen-in fossil magnetostatic field.

 The equation of magneto-solid-mechanics describing coupled oscillations of material displacements ${\bf u}$ and magnetic field $\delta {\bf B}$
 about motionless stationary state of perfectly conducting elastically deformable neutron star matter
 in the presence of the frozen-in fossil magnetic field
 is written as follows \footnote{We remaind that equation for
 $\delta {\bf B}$ in line (\ref{e2.1}) follows from Maxwell equation for Faraday's induction $\delta {\dot {\bf B}}=-c\nabla\times \delta {\bf E}$ where $\delta {\bf E}=-(1/c)[\delta {\bf v}\times {\bf B}]$. This
 link between fluctuating electric field $\delta {\bf E}$ and constant magnetic field ${\bf B}$
 frozen in the conducting flow follows from Ohm's law $\delta {\bf j}/\sigma=\delta {\bf E}+(1/c)[\delta {\bf v}\times {\bf B}]$ corrected for the extremely large, effectively infinite, electrical conductivity, $\sigma\to \infty$, a case of medium designated as a perfect conductor. Taking into account
 that $\delta {\bf v}={\dot {\bf u}}$ and eliminating time derivative from
 obtained in the above manner equation of the field-flow coupling $\delta {\dot {\bf B}}=\nabla\times [\delta {\bf v}\times {\bf B}]$, one arrives at the last equation in the line (\ref{e2.1}).}
  \begin{eqnarray}
 \label{e2.1}
 && \rho{\ddot {\bf u}}=\frac{1}{c}[\delta {\bf j}\times {\bf B}],\quad \delta {\bf j}=\frac{c}{4\pi}[\nabla\times \delta {\bf B}],\quad  \delta {\bf B}=\nabla\times [{\bf u}\times {\bf B}],\\
  \label{e2.2}
 && \rho{\ddot {\bf u}}=\frac{1}{4\pi}[\nabla\times[\nabla\times [{\bf u}\times {\bf B}]]\times {\bf B}],\quad \nabla\cdot  {\bf u}=0,\quad \nabla\cdot  \delta {\bf B}=0.
  \end{eqnarray}
  Scalar  multiplication of (\ref{e2.2}) by ${\dot {\bf u}}$ and integration over the star volume
  leads to the equation of energy balance
   \begin{eqnarray}
 \label{e2.3}
 \frac{\partial }{\partial t}\int \frac{\rho{\dot {\bf u}}^2}{2}\,d{\cal V}=
 \frac{1}{4\pi}\int [\nabla\times[\nabla\times [{\bf u}\times {\bf B}]]\times {\bf B}]\cdot {\dot {\bf u}}\,\,d{\cal V}.
  \end{eqnarray}
 This equation provides a basis for computing the frequency spectrum of nodeless
 vibrations by Rayleigh's
 energy variational method that has been used with success for analysis of shear nodeless vibrations
 driven by solely Hooke's elastic force (Bastrukov et al 2007; 2008).
 The key idea of this method is to use the following separable representation of the field of material displacements
   \begin{eqnarray}
 \label{e2.4}
 {\bf u}({\bf r},t)={\bf a}({\bf r})\,\alpha(t).
   \end{eqnarray}
  In a solid star undergoing global torsional nodeless oscillations about polar axis, the quadrupole and octupole overtones of which are pictured in Fig.1, is given by
  (Bastrukov et al, 2007a)
  \begin{eqnarray}
 \label{e2.5}
 {\dot {\bf u}}({\bf r},t)=[\mbox{\boldmath $\omega$}({\bf r},t)\times {\bf r}],
 \quad \mbox{\boldmath  $\omega$}=A_t\nabla r^\ell\,P_\ell(\cos\theta){\dot\alpha}(t)
  \end{eqnarray}
  so that instantaneous displacements can be written as
   \begin{eqnarray}
 \label{e2.6}
 {\bf a}_t={\cal A}_t\,\nabla \times  [{\bf
 r}\chi({\bf r})],\,\,
 \chi({\bf r})= r^\ell\,P_\ell(\cos\theta):\quad a_r=0,\,\, a_\theta=0,\,\, a_\phi=A_t r^{\ell}(1-\zeta^2)^{1/2}\frac{dP_\ell(\zeta)}{d\zeta}
 \end{eqnarray}
 where the arbitrary constant ${\cal A}_t$ which is eliminated from the boundary condition (Bastrukov et al, 2007a)
 \begin{eqnarray}
 \label{e2.7}
 {\dot {\bf u}}({\bf r},t)=[\mbox{\boldmath {$\omega$}}({\bf r},t)\times {\bf r}]_{r=R}=
 [\mbox{\boldmath $\Omega$}\times {\bf R}],\quad \mbox{\boldmath $\Omega$}={\cal A}_t\nabla P_\ell(\cos\theta){\dot\alpha}(t)\quad\to\quad {\cal A}_t=[R^{(\ell-1)}]^{-1}.
 \end{eqnarray}
  On inserting (\ref{e2.4}) in (\ref{e2.3}) this latter
  equation is reduced to equation for $\alpha(t)$ having the form of equation of harmonic vibrations
   \begin{eqnarray}
 \label{e2.8}
 && \frac{d{\cal H}}{dt}=0,\quad {\cal H}=\frac{{\cal M}{\dot \alpha}^2}{2}+\frac{{\cal K}_m\alpha^2}{2}\quad\to\quad
 {\cal M}{\ddot \alpha}+{\cal K}_m
 {\alpha}=0,
  \end{eqnarray}
  where the integral parameters of inertia ${\cal M}$ and stiffness ${\cal K}_m$ of magnetic Alfv\'en
  vibrations are given by
    \begin{eqnarray}
  \label{e2.9}
 {\cal M}=\int \rho\, {\bf a}^2d{\cal V},\quad {\cal K}_m=\frac{1}{4\pi}\int [\nabla\times[\nabla\times [{\bf a}\times {\bf B}]]\times {\bf B}]\cdot {\bf a}\,d{\cal V}=\frac{1}{4\pi} \int [({\bf B}\cdot \nabla)\, {\bf a}]^2\,d{\cal V}.
  \end{eqnarray}
  In these last equations, the density $\rho$ and frozen in the star magnetostatic fossil magnetic field ${\bf B}$ are considered to be intrinsic characteristics of equilibrium state of the star matter and input parameters
  of the method in question. The mass parameter ${\cal M}$ is positively defined quantity, whereas the sign of
  the stiffness ${\cal K}_m$ depends on specific form of the magnetostatic fossil magnetic field ${\bf B}$
  frozen-in the star. Thus, geometric configuration of the internal field is crucial to the question whether quake-induced shear perturbation resulting in fluctuations of differentially rotational displacements (\ref{e2.10}) are developed as oscillation mode, $\omega^2 > 0$, or relaxation mode, $\omega^2 < 0$, and within the above expounded method this issue, that has been a subject of controversy (Levin 2006, Glampedakis et al 2006; Watts \& Strohmayer 2007), cannot be resolved with no computing ${\cal K}_m$ for each imaginable form of ${\bf B}$.

  The large scale external magnetic fields of radiative magnetospheres of pulsars and magnetars are commonly thought of as produced by the magnetic dipole moment of underlying neutron star and it is generally believed
  that internal magnetic field of the neutron star has a strong poloidal dipolar component.
  With this in mind and for the reason of computational feasibility, in the remainder of  this section we examine the above variational approach by considering an admittedly idealized neutron star model with
  uniform poloidal internal magnetostatic magnetic field ${\bf B}$ directed along the polar axis $z$
  undergoing global torsional oscillations about dipolar magnetic moment axis which are insensitive to the above compositional, core-crust, stratification of the neutron star. The explicit form of spherical components of such a field inside the
 star are given by
  \begin{eqnarray}
  \label{e2.10}
 && B_r=B\zeta,\quad B_\theta=-B(1-\zeta^2)^{1/2},
 \quad B_\phi=0,\quad  r < R, \quad \zeta=
 cos\theta
 \end{eqnarray}
 and components of dipolar configuration outside the star reads
 \begin{eqnarray}
  \label{e2.11}
 && B_r=B\left(\frac{R}{r}\right)^3\zeta,\quad B_\theta=-\frac{B}{2}\left(\frac{R}{r}\right)^3(1-\zeta^2)^{1/2},
 \quad B_\phi=0\quad r > R
 \end{eqnarray}
 where $R$ is the the star radius (e.g. Chandarsekhar \& Fermi 1953).
 Computation of integrals for ${\cal M}$ and ${\cal K}_m$ yields (Bastrukov et al, 1997)
 \begin{eqnarray}
 \label{e2.12}
 {\cal M}(_0t_\ell)=4\pi\rho A_t^2 R^{2\ell+3}\frac{\ell(\ell+1)}{(2\ell+1)(2\ell+3)},\quad
    {\cal K}_m(_0a^t_\ell)=B^2 A_t^2 R^{2\ell +1}\frac{\ell(\ell+1)(\ell^2-1)}{(2\ell+1)(2\ell-1)}
 \end{eqnarray}
 and for the angular frequency of toroidal magnetic Alfv\'en mode of global torsional nodeless oscillations
 we obtain
 \begin{eqnarray}
 \label{e2.13}
 && \omega(_0a^t_\ell)=\omega_A\left[(\ell^2-1)\frac{2\ell+3}{2\ell-1}\right]^{1/2},\quad\quad
 \omega_A=\frac{v_A}{R},\quad v_A=\frac{B}{\sqrt{4\pi\rho}}.
 \end{eqnarray}
 The fundamental frequency of Alfv\'en oscillations $\omega_A$ can be
 conveniently written in the form
 \begin{eqnarray}
 \label{e2.14}
 \omega_A=\sqrt{\frac{B^2R}{3M}},\quad\quad M=\frac{4\pi}{3}\rho\,R^3
 \end{eqnarray}
 where $M$ and $R$ stand for the mass and the radius of star.
 It may be worth noting that normal component of magnetic field under consideration is
 continuous on the surface while the tangential component remains discontinuous and, thus,
 admits, in accord with standard boundary condition of electrodynamics, the surface current, provided that on the surface there is an excess of likely-charged particles distributed with the surface charge density $\sigma$. The considered magnetic field inside the star is identical to that produced by uniformly charged spherical shell of radius $R$ set in the rotation about polar axis with constant angular velocity $\Omega$ and the absolute value of this field
 is given by $B=(2/3)\sigma\Omega R$ (Griffiths 1981). So, the above inferences regarding
 frequency of nodeless Alfv\'en oscillations in the star volume are not affected when
 electrodynamic conditions inside the neutron star model under consideration are compatible with those
 for the latter case of uniformly charged spherical shell.

 In Fig.2, frequencies (left) and periods (right) of the toroidal Alfven mode
 are plotted as functions of multipole degree $\ell$ for a solid star with parameters
 typical for pulsars and magnetars clearly showing general trends of $\nu$ (in Hz) and $P$ (in seconds)
 as a function of multipole degree of oscillations and order of their magnitude.

 In the above computations we have used toroidal (axial) vector field of differentially-rotational nodeless material
 displacements which is one of two fundamental solutions of the vector Laplace equation
\begin{eqnarray}
 \label{e2.15}
 \nabla^2 {\bf a}({\bf r})=0,\quad\quad \nabla\cdot{\bf a}({\bf r})=0
 \end{eqnarray}
 built on fundamental solution of the scalar Laplace equation
 \begin{eqnarray}
 \label{e2.16}
 \nabla^2 \chi({\bf r})=0,\quad\quad \chi({\bf r})=r^{\ell}\,P_\ell(\zeta),\quad\quad \zeta=\cos\theta.
 \end{eqnarray}
 The second fundamental solution
 is given by the even-parity poloidal (polar) vector field (Bastrukov et al 2007a)
 \begin{eqnarray}
 \label{e2.17}
 {\bf a}_p=\frac{A_p}{\ell+1}\nabla\times\nabla\times [{\bf r}\,\chi({\bf r})]=A_p\nabla\,\chi({\bf
 r}),\quad\quad  \chi({\bf r})= r^{\ell}\,P_\ell(\zeta)
 \end{eqnarray}
 which is irrotational: $\nabla\times {\bf a}_p=0$.
The integral parameters of inertia ${\cal M}$ and stiffness ${\cal K}_m$
in the poloidal mode are given by
\begin{eqnarray}
 \label{e2.18}
 {\cal M}(_0a^p_\ell)=4\pi\rho A_p^2 R^{2\ell+1}\frac{\ell}{2\ell+1},\quad
    {\cal K}_m(_0a^p_\ell)=B^2 A_p^2 R^{2\ell -1}\frac{\ell^2(\ell-1)}{2\ell-1}
 \end{eqnarray}
and for the frequency spectrum of poloidal magnetic Alfv\'en mode we get
\begin{eqnarray}
 \label{e2.19}
 \omega(_0a^p_\ell)=\omega^2_A\ell(\ell-1)\frac{2\ell+1}{2\ell-1},\quad\quad
    \omega_A^2=\frac{V_A^2}{R^2}=\frac{B^2}{4\pi\rho R^2}=\frac{B^2R}{3M}.
 \end{eqnarray}
  The spectral formulae like above obtained for toroidal and poloidal Alfven modes\footnote{To the best of our knowledge, classification
  of Alfv\'en vibrational modes as poloidal and toroidal has been introduced
  by Chandrasekhar (1956) in the context of hydromagnetic oscillations of fluid sphere. General
  properties of poloidal and toroidal fields are extensively discussed in (Chandrasekhar 1961; Ferraro \& Plumpton 1961).} are central to theoretical asteroseimology of pulsars and magnetars in the sense they provide the basis for interpretation of observable quasi-periodic oscillations as produced by quake-induced vibrations restored by
  Lorentz force.

   For the purpose of our further consideration we remind that the Lorentz force density, ${\bf f}=(1/c)[{\bf j}\times {\bf B}]$, with the Amp\'ere's current density, ${\bf j}=(c/4\pi)\nabla\times {\bf B}$, having the form
 \begin{eqnarray}
 \label{e2.20}
 && {\bf f}=\frac{1}{4\pi}
 \left[[\nabla\times{\bf B}]\times{\bf B}\right]=
 \frac{1}{4\pi}\left[({\bf B}\cdot \nabla){\bf B}-
 \frac{1}{2}\nabla\, {\bf B}^2\right]
 \end{eqnarray}
  can be represented in terms of Maxwell tensor of magnetic field stresses as (e.g., Mestel 1999)
  \begin{eqnarray}
 \label{e2.21}
  && f_i=\nabla_k T_{ik}, \quad T_{ik}=\frac{1}{8\pi}\left[
 B_i\,B_k+B_k\,B_i - (B_j\,B_j)\delta_{ik}\right].
 \end{eqnarray}
 In an extremely dense matter of solid stars of the finale stage the effects of magnetic buoyancy
 are heavily suppressed and, hence, in the motionless matter of the stationary state of the solid star with
 the above frozen-in uniform magnetic field $f_i=f_i^0(B={\rm const})=0$. However, the Lorentz force surely comes into
 play as a result of quake induced perturbation which leads to
 coupled fluctuations of material displacements ${\bf u}$ and magnetic field $\delta {\bf B}$ in accordance with equation for field-flow coupling $\delta {\bf B}({\bf r},t)=\nabla\times [{\bf u}\times {\bf B}]$. With this in mind, the equation of dynamics (\ref{e2.2}) can be represented in the equivalent tensor form, to wit, in terms of tensor of fluctuating magnetic field stresses (e.g., Franco, Link \& Epstein 2000)
  \begin{eqnarray}
  \label{e2.22}
  \rho{\ddot u}_i=\nabla_k\, \tau_{ik},\quad\quad \tau_{ik} =\frac{1}{4\pi}[B_i\,\delta B_k+B_k\,\delta B_i-(B_j\,\delta B_j)\delta_{ik}],\quad \delta {\bf B}=\nabla\times [{\bf u}\times {\bf B}]
  \end{eqnarray}
  where ${\bf B}$ is the in-advance-given stationary magnetic field. In the text section we show
  that above obtained spectral equation (\ref{e2.13}) for the toroidal Alfv\'en mode of global differentially rotational oscillations of the star matter about homogeneous field frozen-in the star can be regained on the basis
  of this last representation of the Lorentz force density.

  In Fig.3 we plot the frequency and period of Alfv\'enic modes $\omega(_0a_\ell)$ upon multipole degree $\ell$
  of nodeless oscillations of material displacements in a neutron star whose internal magnetic
  field might be approximated by axisymmetric poloidal field of uniform shape. The presented absolute values
  for the frequencies show that they fall in the realm of observable QPOs in flaring SGR's 1806-20 and 1900+14 and, thus, suggesting that Lorentz restoring force must be taken into account when studying global seismic vibrations of magnetars. Together with this it worth emphasizing that obtained one-parametric spectral formula for toroidal Alfv\'en  vibration mode in model under consideration does not match the general trends
  in observed QPOs frequency as a function of multipole degree of torsional vibrations (Watts \& Strohmayer 2007) because the slope of computed frequency $\omega(_0a^t_\ell)$ as a function of $\ell$
  is different from the slope of the overall trends of the observed QPOs frequencies.
  In the next section, we approach to this problem by considering in some details
  the global nodeless torsional oscillations of a neutron star model with the above uniform internal magnetic field under the action of combined forces of shear elastic and magnetic field stresses.

\section{Torsional nodeless vibrations under the joint action of bulk forces of elastic and magnetic field stresses}

 The equation of Newtonian magneto-solid-dynamics appropriate for non-compressional ($\delta \rho=-\rho \nabla\cdot  {\bf u}=0$) shear vibrations of an elastic medium of infinite electrical conductivity are
 \begin{eqnarray}
  \label{e3.1}
  &&\rho {\ddot u}_i=\nabla_k\, \sigma_{ik}+\nabla_k\,\tau_{ik}.
  \end{eqnarray}
  The first term in the right part of (\ref{e3.1}) is the bulk force of elastic stresses obeying the Hooke's law
  \begin{eqnarray}
  \label{e3.2}
  \sigma_{ik}=2\mu\,u_{ik},\quad  u_{ik}=\frac{1}{2}(\nabla_i
  u_k+\nabla_k  u_i),\quad u_{kk}=\nabla_k\,u_k=0
  \end{eqnarray}
  where $\mu$ is the shear modulus of stellar matter linearly relating
  quake-induced shear stresses $\sigma_{ik}$ and resulting shear deformations or strains $u_{ik}$.
  The second term is the above defined Lorentz force represented in terms of fluctuating magnetic field stress
  $\tau_{ik}$. Similar equations have recently been considered in works
  (Piro 2005; Glampedakis, Samuelsson \& Andersson 2006; Lee 2008).
   The conservation of energy is controlled by equation
   \begin{eqnarray}
 \label{e3.4}
 \frac{\partial }{\partial t}\int \frac{\rho{\dot u}_i{\dot u}_i}{2}\,d{\cal V}
 =- \int [\sigma_{ik}+\tau_{ik}]\,{\dot u}_{ik}\,d{\cal V}, \quad {\dot u}_{ik}=\frac{1}{2}[\nabla_i {\dot u}_k+
 \nabla_k {\dot u}_i]
  \end{eqnarray}
  which is obtained after scalar multiplication of (\ref{e3.1}) by ${\dot u}_i$ and integration over the star volume.
  To compute eigenfrequency of torsional nodeless oscillations we again take advantage of the Rayleigh's energy
  method. On inserting separable representation of the field of material displacement
  \begin{eqnarray}
 \label{e3.5}
 u_i({\bf r},t)=a_i({\bf r})\,{\alpha}(t)
 \end{eqnarray}
 in equation of energy balance we again obtain equation of harmonic oscillations
 of temporal amplitude $\alpha(t)$:
 \begin{eqnarray}
\label{e3.6}
 &&  \frac{d{\cal H}}{dt}=0,\quad\quad
 {\cal H}=\frac{{\cal M}{\dot\alpha}^2}{2}+\frac{{\cal K}\alpha^2}{2},\quad\to\quad
 {\cal M}{\ddot \alpha}(t)+{\cal K}\alpha(t)=0,\quad {\cal K}={\cal K}_e+{\cal K}_m
  \end{eqnarray}
  Analytic form for the inertial ${\cal M}$ and for the stiffness ${\cal K}_e$  of shear elastic oscillations
  derived in (Bastrukov et al 2007a; 2007b) are given by
  \begin{eqnarray}
 \label{e3.7}
 && {\cal M}=\int \rho\, a_i({\bf r})\,a_i({\bf r}) d{\cal V},\quad K_e=\frac{1}{2}\int \mu [\nabla_k a_i+
 \nabla_i a_k]\,[\nabla_k a_i+
 \nabla_i a_k]\,d{\cal V}
 \end{eqnarray}
 and for the stiffness of magnetic Alfv\'en shear oscillations we get
 \begin{eqnarray}
 \nonumber
 K_m&=&\frac{1}{8\pi}\int
 \{B_k[\nabla_j(B_j a_i-B_i a_j)]+B_i\,[\nabla_j (B_j a_k-B_k a_j)]\}\,
 (\nabla_k a_i+
 \nabla_i a_k)\,d{\cal V}\\
 \label{e3.8a}
 &=&\frac{1}{4\pi}\int [(B_i\,\nabla_i)a_k]\,[(B_j\,\nabla_j)a_k]\,d{\cal V}.
 \end{eqnarray}
 It should be noted that this last equation can be derived from the standard MHD equations (Chandasekhar 1961).
 Computation of integrals with $B_i$ given by (\ref{e2.3}) and toroidal field
 $a_i$ defined by (\ref{e2.9}) yields
 \begin{eqnarray}
 \label{e3.9}
 && {\cal M}_0(_0t_\ell)=4\pi\rho {\cal A}_t^2 R^{2\ell+3}\frac{\ell(\ell+1)}{(2\ell+1)(2\ell+3)},\\
 \label{e3.10}
 && {\cal K}_e(_0e^t_\ell)=4\pi\mu {\cal A}_t^2 R^{2\ell +1}\frac{\ell(\ell^2-1)}{(2\ell+1)},\quad
 K_m(_0a^t_\ell)=B^2 A_t^2 R^{2\ell +1}\frac{\ell(\ell+1)(\ell^2-1)}{(2\ell+1)(2\ell-1)}
 \end{eqnarray}
  and for the measurable in Hertz total frequency $\nu=\omega/2\pi$ (with
  $\omega=\sqrt{{\cal K}/{\cal M}}$) we obtain
 \begin{eqnarray}
 \label{e3.11}
 && \nu(_0t_\ell)=[\nu_e^2(_0e^t_\ell)+\nu_m^2(_0a^t_\ell)]^{1/2},\\
 \label{e3.12}
 && \nu_e(_0e^t_\ell)=\nu_e\,[(2\ell+3)(\ell-1)]^{1/2},\quad \nu_e=\frac{\omega_e}{2\pi},\quad \omega_e=\frac{c_t}{R},\quad c_t=\sqrt{\frac{\mu}{\rho}},\\
 \label{e3.13}
 && \nu_m(_0a^t_\ell)=\nu_A\,\left[(\ell^2-1)\frac{2\ell+3}{2\ell-1}\right]^{1/2},\,\, \nu_A=\frac{\omega_A}{2\pi},
 \,\,\omega_A=\frac{v_A}{R},\,\, v_A=\frac{B}{\sqrt{4\pi\rho}}.
 \end{eqnarray}
 In Fig.4, the fractional frequency of elastic $\nu_e(_0e^t_\ell)/\nu_e$ oscillations as a function
  of multipole degree $\ell$ is plotted in juxtaposition with
  fractional frequency  $\nu_m(_0a^t_\ell)/\nu_A$ of magnetic Alfv\'en oscillations.
  One sees that the lowest overtones of both elastic and magnetic Alfv\'en modes are of quadrupole degree, $\ell=2$. At $\ell=1$, both parameters of elastic and magneto-mechanical rigidity cancel, ${\cal K}_e(_0e^t_1)=0$ and ${\cal K}_m(_0a^t_1)=0$ and the mass parameter equal to the moment of inertia of rigid sphere, ${\cal M}=(2/5)MR^2$.
  It follows from Hamiltonian that in this dipole case a star sets in rigid body rotation, rather than vibrations, about axis of magnetic dipole moment.

 The obtained spectral equation for total frequency can be conveniently represented in the following form
  \begin{eqnarray}
 \label{e3.14}
 \nu(_0t_\ell)&=&\nu_e[(2\ell+3)(\ell-1)]^{1/2}\,\left[1+\beta\frac{\ell+1}{2\ell-1}\right]^{1/2}
\end{eqnarray}
where
\begin{eqnarray}
 \label{e3.15}
 \nu_e=\sqrt{\frac{\mu}{4\pi^2\rho R^2}},\quad\quad
 \beta=\frac{\omega_A^2}{\omega_e^2}=\frac{\nu_A^2}{\nu_e^2}=\frac{v_A^2}{c_t^2}=\frac{B^2}{4\pi\mu}.
\end{eqnarray}

 In Fig.5 we plot the last two-parametric spectral equation for the total frequency $\nu(_0t_\ell)$ as a function of multipole degree $\ell$ of torsion
 nodeless vibrations computed with indicated values of parameters $\nu_e$ and $\beta$ carrying information
 about mechanical and electrodynamical properties of the neutron star matter which are
 adjusted so as to reproduce the observable frequency of QPOs (symbols) during the flare of SGRs  1806-20 and 1900+14;
 the data from (Watts \& Strohmayer 2007, Samuelsson \& Andersson 2007).
 This figure demonstrates that the detected QPOs can be consistently explained from the viewpoint of the
 the considered model as produced by global torsional nodeless vibrations when and only when combined forces of shear elastic and magnetic field stresses come into play in a coherent fashion.
 On the other hand, in our previous study reported in recent paper (Bastrukov et al, 2008a) it has been shown that this set of QPOs data can be properly described, with the same degree of accuracy,
 on the basis of two-parametric spectral formula that has been derived on the basis of
 a two-component, core-crust, model of quaking neutron star presuming that detected QPOs are produced by axisymmetric torsional nodeless seismic vibrations driven by a solely elastic restoring force and locked in the peripheral finite-depth seismogenic layer.
 The truth is, most probably, somewhere in between and in order to attain more definite conclusions further investigations,  both theoretical and observational, are needed. From a computational argument, the nodeless torsional oscillations entrapped in the neutron star crust as well as in the star models with non-uniform axisymmetric internal magnetic field and non-homogeneous profile of shear modulus, requires a more elaborate mathematical treatment. To avoid destructing attention
 from the newly obtained results presented  here [some of which are of interest, as is hoped, for general theoretical seismology (Lay \& Wallace 1995; Aki \& Richards 2002)], we postpone a discussion of these latter cases to a
 forthcoming article.

 A special comment should be made regarding the link between neutron stars and atomic nuclei
 which can be considered to be similar objects as far as mechanical properties of degenerate nucleon material
 of normal nuclear density are concerned.
 It follows from the nuclear solid-globe model
 of giant resonances (within the framework of which these fundamental modes of nuclear excitations
 are properly described in term of shear elastic vibrations of an ultra small spherical piece
 of nucleon Fermi-solid (regarded as continuous matter) that the shear modulus
 is given by  $\mu\sim 10^{33}$ dyn cm$^{-2}$ (e.g., Bastrukov et al, 2008b). Making use of this value of $\mu$ in
 the expression for parameter $\beta=[B^2/4\pi\mu]\sim 0.5-0.6$ entering the above derived spectral
 equation for QPOs frequency, we can get an independent estimate for the intensity of the magnetic field frozen in
 the star. With the above values of $\beta$ and $\mu$ one finds that $B$
 falls in the range $10^{15}< B < 10^{16}$ Gauss, that is, in the realm typical of magnetar magnetic fields.
 This latter inference demonstrates the theoretical potential of the asteroseimology showing how
 physical interpretation of the QPOs as produced by quake induced torsional vibrations of neutron
 star can be used to extract information about properties of the neutron star matter.

\section{Summary}

 There is a common belief today that gross features of the asteroseimology of pulsars and magnetars
 can be understood on the basis of a solid star model presuming that quake-induced shear vibrations restored by
 bulk forces of intrinsic stresses of different physical nature are governed by described by solid mechanics or elastodynamics  (e.g., Hansen \& van Horn 1979; McDermott, van Horn \& Hansen 1988; Bastrukov,
 Weber \& Podgainy 1999; Bastrukov et al 2007a). This point of view is quite different from theoretical approach to the asteroseismology of the main sequence stars at the base of which lies the liquid star model whose vibrations are treated within the framework of fluid-mechanical theory of continuous media, as is the case of helioseismology (e.g., Christensen-Dalsgaard 2002).

 The main purpose of this work was  to examine the magneto-solid-mechanical
 variational method of the asteroseismology of neutron star by probing its interior with non-radial global differentially rotational, torsional, vibrations with nodeless toroidal field of material displacements,
 which are insensitive to compositional stratification of the star matter.  Bearing in mind that external magnetic fields of pulsars and magnetars are commonly thought of as produced by magnetic dipole moment of underlying neutron star we have considered a model of a neutron star with perhaps simplest, from the viewpoint of computational feasibility, imaginable configuration of the magnetostatic fossil magnetic field, pictured in Fig.1.
 Proceeding from this admittedly idealized model, the two-parametric spectral equation for the frequency of global torsional vibrations has been derived in analytic form showing the larger multipole degree $\ell$ of torsional nodeless vibrations the higher is the frequency $\omega(_0t_\ell)$.
 The application of the obtained two-parametric spectral formula to modal analysis of QPOs during the flare of SGR 1806-20 and SGR 1900+14 shows that data on the QPOs frequencies with $\ell$ from interval $2\leq \ell\leq 20$ can be consistently interpreted as produced by global torsional nodeless vibrations restored by combined forces of shear elastic and magnetic field stresses. This inference is, of course, suggestive rather than conclusive, in view of adopted a highly idealized configuration of fossil magnetostatic field frozen in the star,
 and too much remains to be done to be at all certain of interpretations suggested for QPOs in the X-ray flux during the giant flares of the above magnetars.

\acknowledgments
 The authors are grateful to Dr. Judith Bunder (UNSW, Sydney) for critical reading and
 suggestions regarding the text of paper.
 This work is partly supported by NSC of Taiwan,
 under grants  NSC-96-2628-M-007-012-MY3 and NSC-97-2811-M-007-003.

\clearpage

\begin{figure}[h]
\centering\
\includegraphics[width=12.0cm]{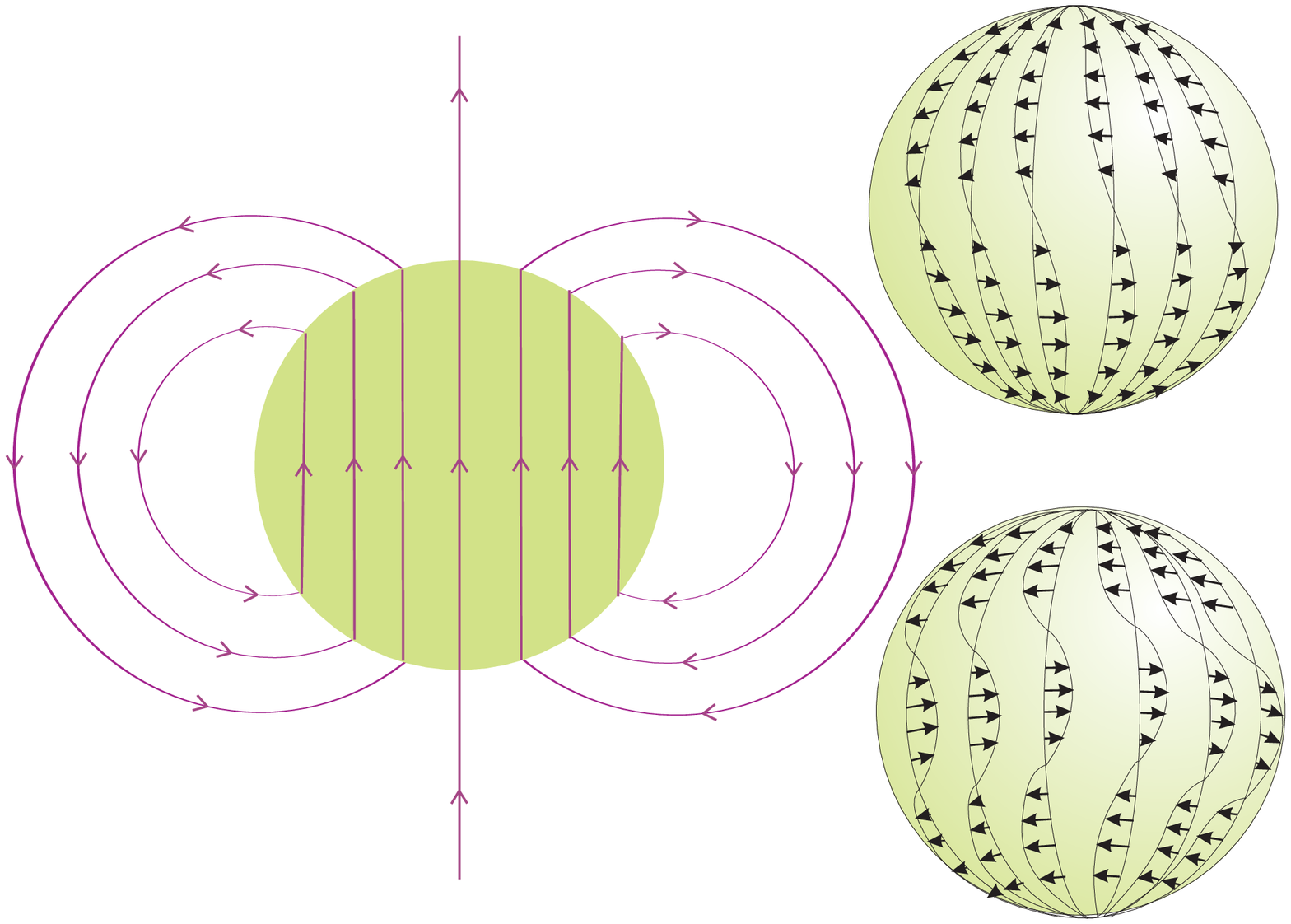}
\caption{\small
Fiducial model of a solid star with uniform magnetic field inside and dipole outside undergoing
global torsional nodeless oscillations
in quadruple and octupole overtones.}
\end{figure}

\begin{figure}[h]
\centering\
\hspace*{-2.cm}
\includegraphics[width=16.0cm]{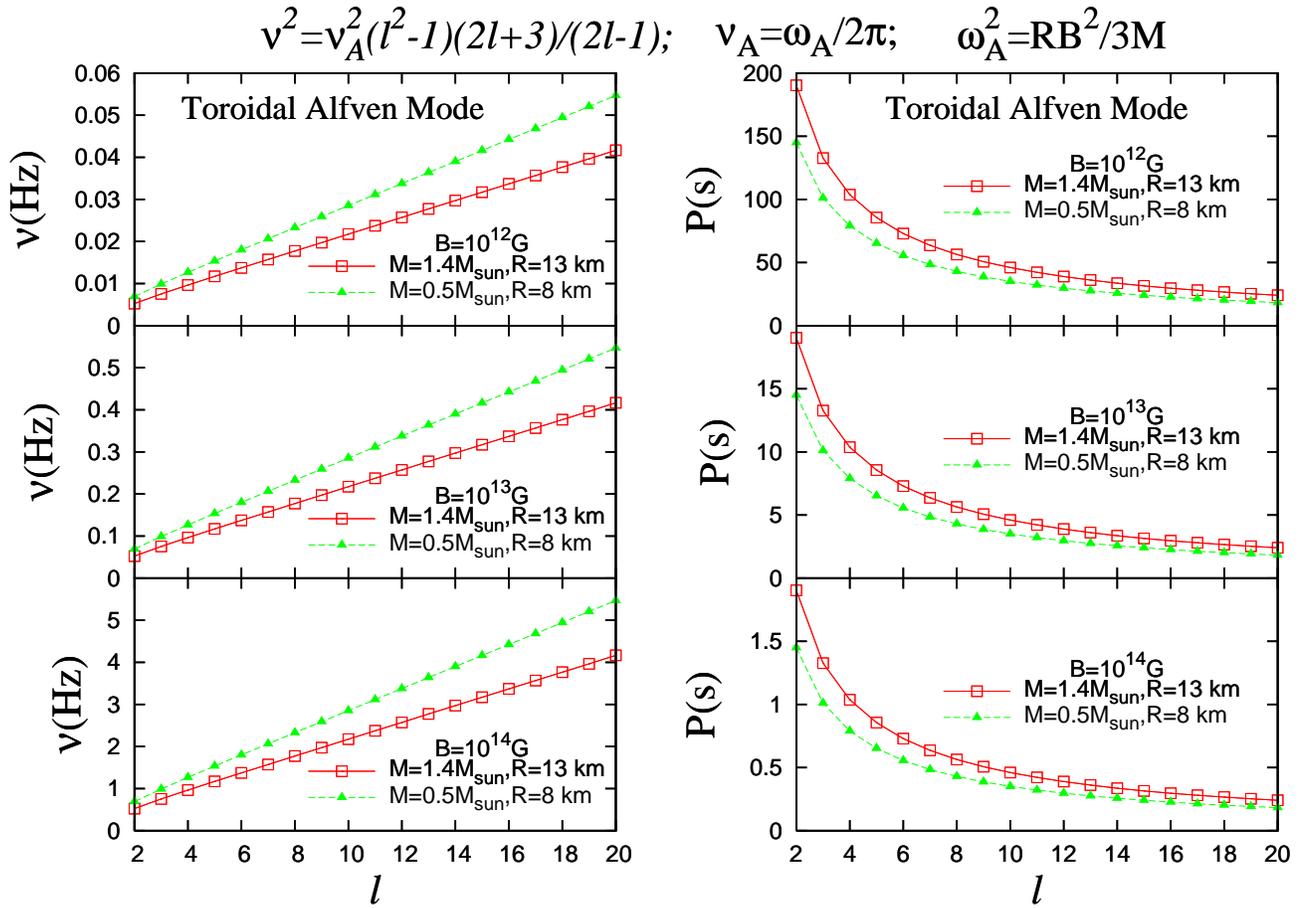}
\vspace*{0.5cm}
\caption{\small Frequencies (left) and periods (right) of toroidal magnetic Alfven vibrational mode as functions of multipole degree $\ell$ in the neutron star models with pointed out intensities of uniform magnetic field inside and dipole outside.}
\end{figure}

\begin{figure}[h]
\centering\
\includegraphics[width=16.0cm]{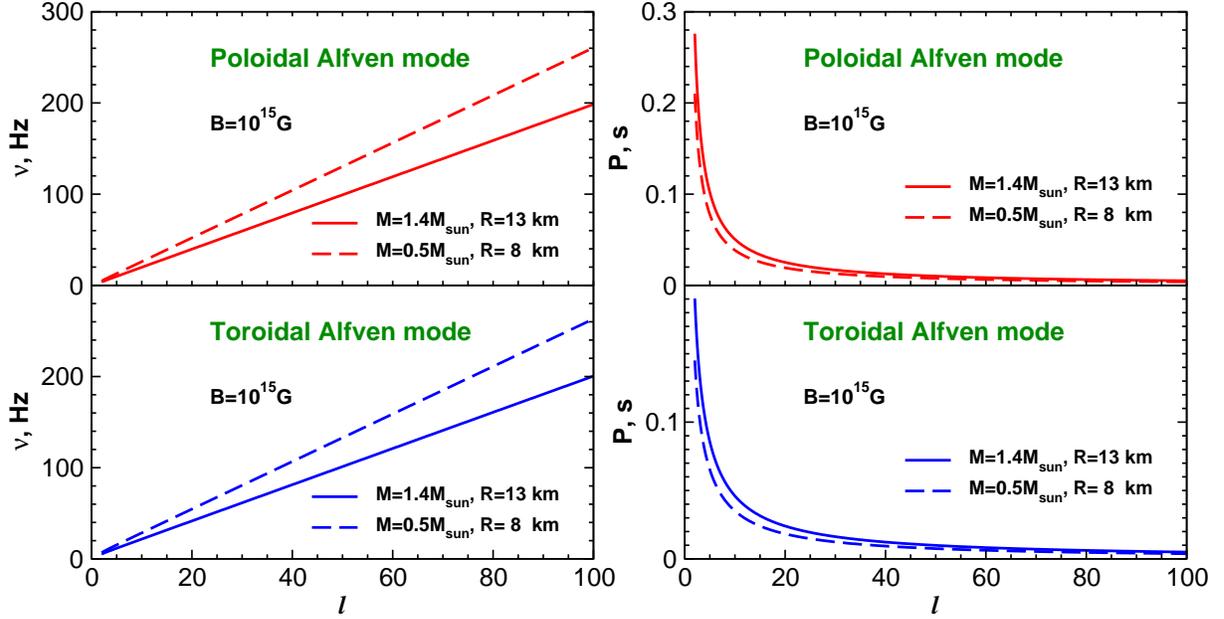}
\caption{\small Computed frequencies and periods of poloidal and toroidal Alfv\'en modes in the neutron star
models undergoing global spheroidal and
torsional shear nodeless vibrations, respectively, driven by Lorentz restoring force.}
\end{figure}

 \begin{figure}[h]
\centering\
\includegraphics[width=12.0cm]{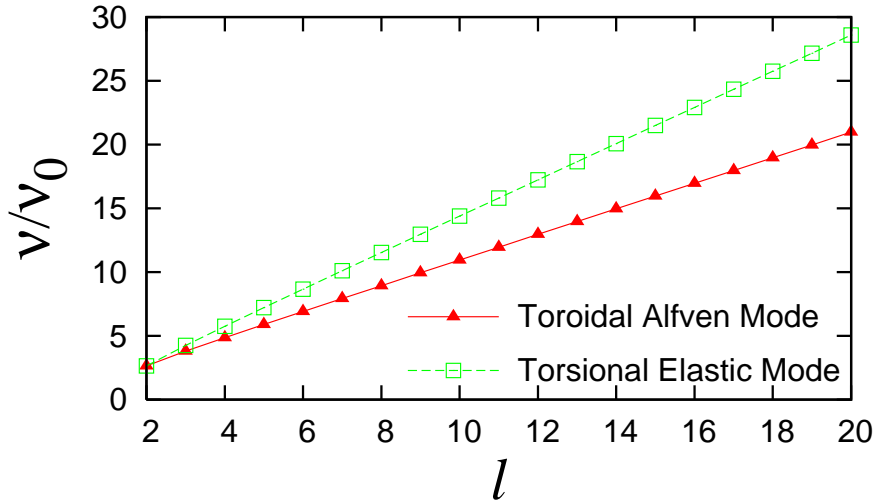}
\caption{\small
 Fractional frequencies $\nu/\nu_0$ of toroidal Alfven mode (with $\nu_0=\nu_A$) and torsional elastic mode (with $\nu_0=\nu_e$) in a solid star undergoing differentially
rotational nodeless oscillations about polar axis of uniform inside and dipole outside magnetic field.
 }
\end{figure}

\begin{figure}[h]
\centering\
\hspace*{3.cm}
\includegraphics[width=16.0cm]{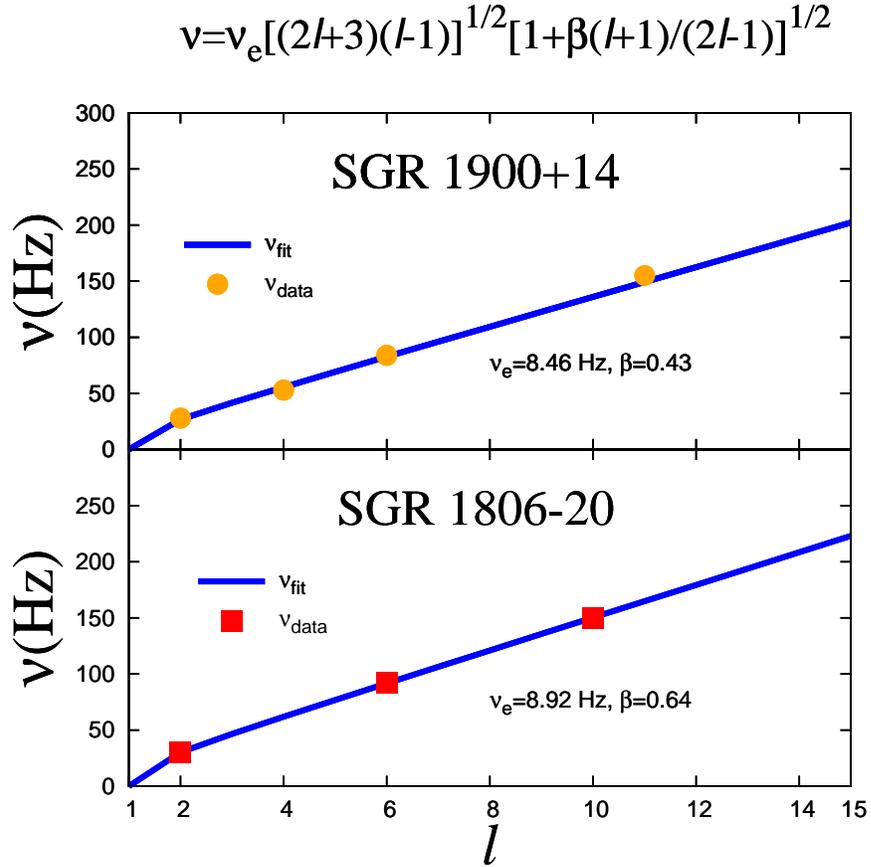}
\vspace*{0.5cm}
\caption{\small
 Theoretical fit of data (symbols) on QPOs frequencies during the flare of SGRs 1806-20 and 1900+14
 by the above spectral equation for the frequency of
 axisymmetric torsional nodeless vibrations about axis of dipole magnetic moment of magnetar under the joint action of bulk forces of elastic and magnetic stresses.}
\end{figure}

\end{document}